\documentclass[conference]{IEEEtran}
\IEEEoverridecommandlockouts

\usepackage{subcaption}

\usepackage{cite}
\usepackage{amsmath,amssymb,amsfonts}
\usepackage{graphicx}
\usepackage{textcomp}
\usepackage{xcolor}
\def\BibTeX{{\rm B\kern-.05em{\sc i\kern-.025em b}\kern-.08em
T\kern-.1667em\lower.7ex\hbox{E}\kern-.125emX}}
\usepackage{adjustbox}
\newlength\mylen
\usepackage{algorithm}
\usepackage{algpseudocode}
\usepackage{tabularx}
\usepackage{tikz}
\usepackage{array}
\newcolumntype{C}{>{\hfil$}p{\mylen}<{$\hfil}}
\usetikzlibrary{matrix}

\usepackage[switch]{lineno}

\usepackage[round-mode=places,round-precision=0,group-separator={,},output-decimal-marker={.}]{siunitx}

\begin{document}

%
%
\newcommand{\DO}{{\textbf{do}}}
\newcommand{\ELSE}{{\textbf{else}}}
\newcommand{\ELSEIF}{{\textbf{else if}}}
\newcommand{\ENDFOR}{{\textbf{end for}}}
\newcommand{\ENDIF}{{\textbf{end if}}}
\newcommand{\FOR}{{\textbf{for}}}
\newcommand{\IF}{{\textbf{if}}}
\newcommand{\STEP}{{\textbf{step}}}
\newcommand{\SUCHTHAT}{{\textbf{such that}}}
\newcommand{\THEN}{{\textbf{then}}}
\newcommand{\TO}{{\textbf{to}}}
\newcommand{\adj}[2]{\ensuremath{\mbox{adj}_{#1}(#2)}}
\newcommand{\bc}[1]{\ensuremath{\mbox{bc}(#1)}}
\newcommand{\cdiv}[1]{\ensuremath{\mbox{cdiv}(#1)}}
\newcommand{\chain}[1]{\ensuremath{\mbox{chain}(#1)}}
\newcommand{\cmod}[1]{\ensuremath{\mbox{cmod}(#1)}}
\newcommand{\Comp}[1]{\ensuremath{\mbox{Comp}(#1)}}
\newcommand{\depth}[1]{\ensuremath{\mbox{depth}(#1)}}
\newcommand{\diffsiz}{\ensuremath{\mbox{diff\_siz}}}
\newcommand{\Edge}[2]{\ensuremath{\{#1,#2\}}}
\newcommand{\Efill}{\ensuremath{E^{+}}}
\newcommand{\False}{{\textbf{false}}}
\newcommand{\fchild}[1]{\ensuremath{\mbox{fchild}(#1)}}
\newcommand{\find}[1]{\ensuremath{\mbox{FIND}(#1)}}
\newcommand{\Free}{\ensuremath{\mbox{FREE}}}
\newcommand{\Gfill}{\ensuremath{G^{+}}}
\newcommand{\glbind}[1]{\ensuremath{\mbox{glbind}(#1)}}
\newcommand{\Gskel}{\ensuremath{G^{-}}}
\newcommand{\hadj}[2]{\ensuremath{\mbox{hadj}_{#1}(#2)}}
\newcommand{\ijsup}{\ensuremath{\mbox{ijsup}}}
\newcommand{\imax}{\ensuremath{i_{\mbox{\scriptsize max}}}}
\newcommand{\imaxx}{\ensuremath{i_{\mbox{\tiny max}}}}
\newcommand{\imin}{\ensuremath{i_{\mbox{\scriptsize min}}}}
\newcommand{\iminn}{\ensuremath{i_{\mbox{\tiny min}}}}
\newcommand{\indmap}[1]{\ensuremath{\mbox{indmap}(#1)}}
\newcommand{\indmapn}{$\mbox{indmap}(1$:$n)$}
\newcommand{\Int}[1]{\ensuremath{\mbox{int\_siz}(#1)}}
\newcommand{\jlist}{\ensuremath{\mbox{jlist}}}
\newcommand{\jmax}{\ensuremath{j_{\mbox{\scriptsize max}}}}
\newcommand{\jmin}{\ensuremath{j_{\mbox{\scriptsize min}}}}
\newcommand{\kmax}{\ensuremath{k_{\mbox{\scriptsize max}}}}
\newcommand{\kmin}{\ensuremath{k_{\mbox{\scriptsize min}}}}
\newcommand{\ladj}[2]{\ensuremath{\mbox{ladj}_{#1}(#2)}}
\newcommand{\lca}[1]{\ensuremath{\mbox{lca}(#1)}}
\newcommand{\level}[1]{\ensuremath{\mbox{level}(#1)}}
\newcommand{\Lnz}[1]{\ensuremath{\mbox{Lnz}(#1)}}
\newcommand{\nul}{\ensuremath{\mbox{\bf null}}}
\newcommand{\parent}[1]{\ensuremath{\mbox{parent}(#1)}}
\newcommand{\prevlf}[1]{\ensuremath{\mbox{prevlf}(#1)}}
\newcommand{\relind}[1]{\ensuremath{\mbox{relind}(#1)}}
\newcommand{\relindB}[1]{\ensuremath{\mbox{relindB}(#1)}}
\newcommand{\rep}[1]{\ensuremath{\mbox{rep}(#1)}}
\newcommand{\repn}{$\mbox{rep}(1$:$n)$}
\newcommand{\Right}{\ensuremath{\mbox{Right}}}
\newcommand{\set}[1]{\ensuremath{\mbox{set}(#1)}}
\newcommand{\setsize}{\ensuremath{\mbox{set\_size}}}
\newcommand{\snode}[1]{\ensuremath{\mbox{snode}(#1)}}
\newcommand{\Sup}[1]{\ensuremath{\mbox{sup}(#1)}}
\newcommand{\supi}{\ensuremath{\mbox{supi}}}
\newcommand{\supj}{\ensuremath{\mbox{supj}}}
\newcommand{\tn}{$t(1$:$n)$}
\newcommand{\Tree}[1]{\ensuremath{T_{r}[#1]}}
\newcommand{\True}{{\textbf{true}}}
\newcommand{\Tsize}[1]{\ensuremath{\mbox{Tsize}(#1)}}
\newcommand{\union}[1]{\ensuremath{\mbox{UNION}(#1)}}

\makeatletter
\algnewcommand{\LineComment}[1]{\Statex \hskip\ALG@thistlm /*\ #1\ */}
\newcommand{\multiline}[1]{%
\begin{tabularx}{\dimexpr\linewidth-\ALG@thistlm}[t]{@{}X@{}}
#1
\end{tabularx}
}
\makeatother
\algnewcommand{\IIf}[1]{\State\algorithmicif\ #1\ \algorithmicthen}
\algnewcommand{\EndIIf}{\unskip\ \algorithmicend\ \algorithmicif}
\algnewcommand{\FFor}[1]{\State\algorithmicfor\ #1\ \algorithmicdo}
\algnewcommand{\EndFFor}{\unskip\ \algorithmicend\ \algorithmicfor}
\algrenewcommand\algorithmicindent{1.0em}%
\algnewcommand{\IfThenElse}[3]{
\State \algorithmicif\ #1\ \algorithmicthen\ #2\ \algorithmicelse\ #3}

\title{GPU Accelerated Sparse Cholesky Factorization*\vspace{-1.5ex}\\
\thanks{
This work was supported in part by the U.S. Department of Energy, Office of Science, Office of Advanced Scientific Computing Research and Office of Basic Energy Sciences, Scientific Discovery through Advanced Computing (SciDAC) Program through the FASTMath Institute and BES Partnership under Contract No. DE-AC02-05CH11231 at Lawrence Berkeley National Laboratory.
We used resources at the DOE NERSC facility for the experiments.
}
}

\author{\IEEEauthorblockN{M. Ozan Karsavuran}
\IEEEauthorblockA{\textit{Lawrence Berkeley National Laboratory}\\
California, USA \\
mokarsavuran@lbl.gov}
\and
\IEEEauthorblockN{Esmond G. Ng}
\IEEEauthorblockA{\textit{Lawrence Berkeley National Laboratory}\\
California, USA \\
egng@lbl.gov}
\and
\IEEEauthorblockN{Barry W. Peyton}
\IEEEauthorblockA{\textit{Dalton State College}\\
Georgia, USA \\
bpeyton@daltonstate.edu}
}

\maketitle

\begin{abstract}
The solution of sparse symmetric positive definite linear systems is an important computational kernel in large-scale scientific and engineering modeling and simulation. We will solve the linear systems using a direct method, in which a Cholesky factorization of the coefficient matrix is performed using a right-looking approach and the resulting triangular factors are used to compute the solution. Sparse Cholesky factorization is compute intensive. In this work we investigate techniques for reducing the factorization time in sparse Cholesky factorization by offloading some of the dense matrix operations on a GPU.
We will describe the techniques we have considered. We achieved up to 4x speedup compared to the CPU-only version.
\end{abstract}

\begin{IEEEkeywords}
sparse matrices,
right-looking Cholesky factorization,
GPU acceleration,
supernodes
\end{IEEEkeywords}

%
%
\section{Introduction}
\label{sec:intro}

Cholesky factorization is a common way to factorize the coefficient matrix into 
triangular factors, which are used to compute the solution of a linear system.
Today all sparse Cholesky factorization algorithms utilize supernode structure.
A supernode is a set of columns of the factor matrix that have the same sparsity structure.
Hence, supernodes form dense submatrices.
One can use BLAS routines on these dense submatrices in order to achieve high performance.

Our primary focus in this paper is accelerating serial sparse Cholesky factorization by offloading some BLAS computations to a GPU.
That is, we do not propose a new parallel algorithm, rather we utilize existing efficient libraries that run in parallel on GPU.
In this way, we show that it is relatively easy to obtain reasonable speedups from a serial implementation.
On the other hand, in our methods most of the floating-point computations of the factorization are performed on GPU using multiple threads.

Sparse Cholesky algorithms have been widely studied in the literature. 
In right-looking methods~(RL)~\cite{factor}, updates from the current supernode, which is factorized, are applied to supernodes to the right.
In~\cite{factor} two new RL variants are introduced, and it is shown that they are superior to or competitive with other methods in terms of both time and storage requirements.
Hence, in this work we will focus on these two new RL variants.

\section{Two recent factorization variants}
\label{sec:rlalgs}
\vspace{-1ex}
Here we will briefly describe both our notation and two recent right-looking supernodal sparse Cholesky algorithms, namely RL and RLB (see later), introduced in~\cite{factor}.
In this paper, we will explain these algorithms mainly by using examples.
We refer the reader to~\cite{factor} for full details.
These algorithms will be our base algorithms in which we will offload some computations to GPU.

Consider the Cholesky factorization $A=LL^T$, where $A$ is an $n\times n$ sparse symmetric positive definite matrix and $L$ is a lower triangular matrix. 
$A_{*,j}$ will denote column~$j$ of matrix~$A$,
and $A_{*,J}$ will denote a set of contiguous columns of matrix $A$, where $J$ denotes the set of column indices.


The elimination tree is based on the sparsity structure of the factor matrix~\cite{Liu90}; it is constructed as follows.
Each column~$j$ of the factor matrix is represented by a vertex~$j$.
Then the parent of vertex~$j$ is the vertex with minimum row index~$i$ among the nonzero values in column~$j$ except $j$.
The set of row indices of column~$j$'s nonzeros (except $j$) is a subset of the row indices of the parent column of~$j$'s nonzeros~\cite{Liu90,Schreiber82}.




In order to efficiently perform updates from a supernode to another supernode, 
we need to match their indices.
A way of doing this is called relative indices, introduced by Schreiber~\cite{Schreiber82} and used by Ashcraft~\cite{Ashcraft87} in his multifrontal implementation.

Assume that $J'$ is an ancestor of~$J$ in the supernodal elimination tree.
The relative indices $\relind{J,J'}$ contain an index for each global index~$i$ in the intersection set of the row indices of $J$ and the row indices of $J'$ such that it is the distance of $i$ from the bottom of the set of the indices of $J'$.
These indices are essential for efficiently applying updates from one supernode to another.

\begin{figure}[htp]
\centering
\begin{subfigure}[b]{.5\linewidth}
\scriptsize
\setlength{\arraycolsep}{1.2pt}
\begin{center}
\begin{eqnarray*}
{\color{white} L} & {\color{white} =} &
{\color{white} \left[
\begin{array}{cc|cc|ccc|cc|cc|cccc}
\multicolumn{2}{c}{\color{black} J_1} &
\multicolumn{2}{c}{\color{black} J_2} &
\multicolumn{3}{c}{\color{black} J_3} &
\multicolumn{2}{c}{\color{black} J_4} &
\multicolumn{2}{c}{\color{black} J_5} &
\multicolumn{4}{c}{\color{black} J_6} \\
\vspace{-3ex}

{\color{white} 1} & 
{\color{white} 2} & 
{\color{white} 3} & 
{\color{white} 4} & 
{\color{white} 5} & 
{\color{white} 6} & 
{\color{white} 7} & 
{\color{white} 8} &
{\color{white} 9} & 
{\color{white} 10} & 
{\color{white} 11} & 
{\color{white} 12} & 
{\color{white} 13} & 
{\color{white} 14} & 
{\color{white} 15} 
\end{array}
\right]} \\
L & = &
\left[
\begin{array}{cc|cc|ccc|cc|cc|cccc}
{1} & \multicolumn{14}{c}{} \\
\ast & {2} & \multicolumn{13}{c}{} \\
& & 3 & \multicolumn{12}{c}{} \\
&&\ast& 4 & \multicolumn{11}{c}{} \\
&&& & 5 & \multicolumn{10}{c}{} \\
\ast&\ast&& & \ast& 6 & \multicolumn{9}{c}{} \\
\ast&\ast&& & \ast&\ast & 7 & \multicolumn{8}{c}{} \\
& &\ast &\ast&&&& 8 & \multicolumn{7}{c}{} \\
& & \ast&\ast&&&&\ast & 9 & \multicolumn{6}{c}{} \\
& & &&&&& & &10&\multicolumn{5}{c}{} \\
&&&&&&&&&\ast&11&\multicolumn{4}{c}{} \\
&&&&&&&\ast&\ast&& &12&\multicolumn{3}{c}{} \\
& & \ast& \ast& \ast& \ast& \ast& \ast&\ast & &&\ast&13&\multicolumn{2}{c}{} \\
\ast&\ast&&&\ast&\ast&\ast& & &\ast &\ast&\ast& \ast&14&\multicolumn{1}{c}{} \\
& & & &\ast &\ast &\ast & & &\ast &\ast&\ast& \ast& \ast&15
\end{array}
\right]
\end{eqnarray*}
\end{center}
\vspace{-3ex} 
\label{fig:supernode2}
\end{subfigure}%
\begin{subfigure}[b]{0.5\linewidth}
\raggedleft
\scriptsize
\setlength{\arraycolsep}{2pt}

\mbox{}

\begin{tikzpicture}[scale=.46]

%
%
\node (1) at (-2.5,2) [circle, draw] {$J1$};
\node (3) at (-2.5,4) [circle, draw] {$J3$};
\node (2) at (-0.5,2) [circle, draw] {$J2$};
\node (4) at (-0.5,4) [circle, draw] {$J4$};

\node (5) at (1.5,4) [circle, draw] {$J5$};
\node (6) at (-0.5,6.5) [circle, draw] {$J6$};
%
\draw (1) -- (3);
\draw (2) -- (4);
\draw (3) -- (6);
\draw (4) -- (6);
\draw (5) -- (6);
\end{tikzpicture}
\vspace{5ex}
\label{fig:set}
\end{subfigure}
\vspace{-2ex}
\caption{The supernodes (left) and supernodal elimination tree (right) of a sparse Cholesky factor $L$.}
\label{fig:supernode2X}
\vspace{-5ex}
\end{figure}
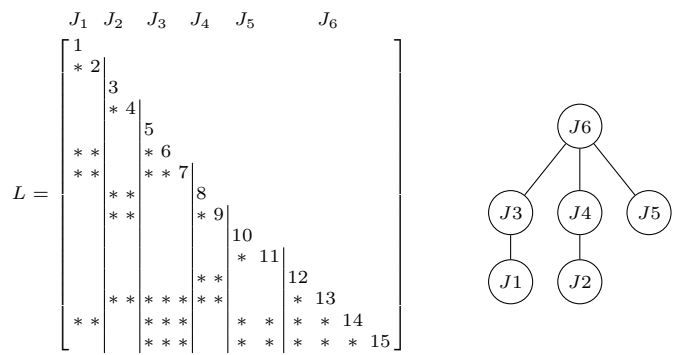


An example $15\times 15$ factor matrix~$L$ and the corresponding supernodal elimination tree are shown on the left and the right, respectively, in Figure~\ref{fig:supernode2X}.
As seen in the figures, supernode~$J_1$ contains columns 1 and 2. That is, $J_1=\{1,2\}$. 
Similarly, $J_2=\{3,4\}$ and so on.
As seen in the figures, supernode~$J_1$ updates supernodes~$J_3$ and $J_6$, whereas supernode~$J_2$ updates supernodes~$J_4$ and $J_6$. 
Supernode~$J_5$ also updates supernode~$J_6$, but it is not updated by any other supernode.

\subsection{A right-looking method (RL)}
\label{sec:RL}

The algorithm processes supernodes starting from the leftmost one.
When the algorithm starts the computation of the current supernode~$J$, all updates from supernodes to the left have already been applied to supernode~$J$.
RL first invokes DPOTRF (for computing a dense Cholesky factorization) on the dense lower triangular part of the supernode and then DTRSM (for dense triangular solution) on the rectangular part of the supernode.
Thus, the current supernode is factorized. 
Note that a supernode is stored in a dense array.
For example, supernode~$J_1$ is stored in an array of size $5\times2$, and supernode~$J_3$ is stored in an array of size $6\times3$.
BLAS operations are performed on these arrays.

After factorizing the current supernode, the corresponding update matrix is computed using a DSYRK (for symmetric rank-k update) BLAS call.
Note that this algorithm requires temporary working storage for the update matrices.
The temporary working storage is preallocated so that it can store the largest update matrix during the factorization.

\begin{figure}[htp]
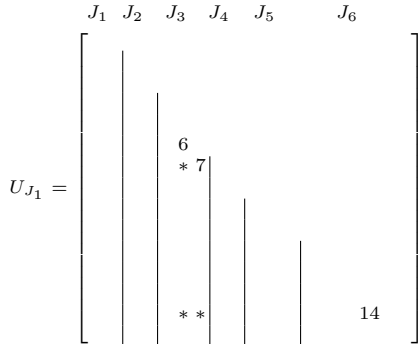

\vspace{-4ex}
\scriptsize
\setlength{\arraycolsep}{1.2pt}
\begin{center}
\begin{eqnarray*}
{\color{white} L} & {\color{white} =} &
{\color{white} \left[
\begin{array}{cc|cc|ccc|cc|cc|cccc}
\multicolumn{2}{c}{\color{black} J_1} &
\multicolumn{2}{c}{\color{black} J_2} &
\multicolumn{3}{c}{\color{black} J_3} &
\multicolumn{2}{c}{\color{black} J_4} &
\multicolumn{2}{c}{\color{black} J_5} &
\multicolumn{4}{c}{\color{black} J_6} \\
\vspace{-3ex}
{\color{white} 1} & 
{\color{white} 2} & 
{\color{white} 3} & 
{\color{white} 4} & 
{\color{white} 5} & 
{\color{white} 6} & 
{\color{white} 7} & 
{\color{white} 8} &
{\color{white} 9} & 
{\color{white} 10} & 
{\color{white} 11} & 
{\color{white} 12} & 
{\color{white} 13} & 
{\color{white} 14} & 
{\color{white} 15} 
\end{array}
\right]} \\
U_{J_1} & = &
\left[
\begin{array}{cc|cc|ccc|cc|cc|cccc}
{\color{white} 1} & \multicolumn{14}{c}{} \\
& {\color{white} 2} & \multicolumn{13}{c}{} \\
& & {\color{white} 3}& \multicolumn{12}{c}{} \\
&&& {\color{white} 4}& \multicolumn{11}{c}{} \\
&&& & {\color{white} 5}& \multicolumn{10}{c}{} \\
&&& & & 6 & \multicolumn{9}{c}{} \\
&&& & & \ast& 7 & \multicolumn{8}{c}{} \\
& & &&&&& {\color{white} 8}& \multicolumn{7}{c}{} \\
& & &&&&& & {\color{white} 9}& \multicolumn{6}{c}{} \\
& & &&&&& & &{\color{white} 10}&\multicolumn{5}{c}{} \\
&&&&&&&&&&{\color{white} 11}&\multicolumn{4}{c}{} \\
&&&&&&&&&& &{\color{white} 12}&\multicolumn{3}{c}{} \\
& & & & & & & & & &&&{ {\color{white} 13}}&\multicolumn{2}{c}{} \\
&&&&&\ast&\ast& & & &&& &14&\multicolumn{1}{c}{} \\
& & & & & & & & & &&& & &{\color{white} 15}
\end{array}
\right]
\end{eqnarray*}
\vspace{-4ex}
\caption{The update matrix computed from the supernode~$J_1$ shown in Figure~\ref{fig:supernode2X}.}
\label{fig:updatematrix2}
\end{center}
\vspace{-5ex}
\end{figure}

Figure~\ref{fig:updatematrix2} displays the update matrix computed by supernode~$J_1$.
Since supernode~$J_1$ updates supernodes~$J_3$ and $J_6$, the update matrix has nonzeros that lie within these supernodes.

Then the update matrix will be assembled into factor storage for each ancestor of the current supernode~$J$.
Doing this efficiently requires utilizing the relative indices effectively.
Assume that the current supernode is $J_1$.
Then we know that 
\[
\relind{J_1,J_3} = \left[
\scriptsize
 \begin{array}{c}
9 \\ 8 \\ 1
\end{array}
\right]
\mbox{and }
\relind{J_3,J_6} = \left[ 
\scriptsize
\begin{array}{c}
2 \\ 1 \\ 0 \\
\end{array}
\right].
\]
Therefore, the three rows in the update matrix’s columns 6 and 7 will be assembled to supernode~$J_3$'s relative indices $\{1,8,9\}$.
Since supernode~$J_1$ is updating supernode~$J_6$ as well, we also need $\relind{J_1,J_6}$. 
Observe that 
$
\relind{J_1,J_6} = \left[ 
\scriptsize
\begin{array}{c}
1
\end{array}
\right].
$
Generalized relative indices open the way for an efficient right-looking supernodal sparse Cholesky
algorithm~\cite{factor}.

%
%

\subsection{A right-looking blocked method (RLB)}
\label{sec:RLB}

The first stage of RLB is the same as RL’s first stage.
That is, RLB first factorizes the current supernode~$J$ with level-3 BLAS operations, namely DPOTRF and DTRSM.
Then instead of computing the whole update matrix of supernode~$J$ at once, RLB processes 
blocks of the supernode~$J$ one by one,  where each block contains the maximum number of consecutive dense rows in~$J$.
That is, RLB decomposes the update process into many DSYRK and DGEMM (for dense matrix-matrix product) calls.
The advantage of this decomposition is that RLB does not require a temporary update matrix; it directly updates the ancestor supernodes within factor storage.
Also, the indexing is simpler; RLB just requires one generalized relative index for the each block in the supernode, whereas RL requires generalized relative indices for each row in the supernode.

RLB processes each pair of blocks~$B$ and~$B'$ in supernode~$J$, where $B$ is above $B'$.
Assume that $B \subseteq P$, where $P$ is an ancestor of $J$.
%
%
For example, again consider supernode~$J_1$ displayed in Figure~\ref{fig:supernode2X}.
We have two blocks, let's say $B=\{6,7\}$ and $B'=\{14\}$.
Then the algorithm first updates the diagonal part~$L_{B,B}$ of supernode~$J_3$ by invoking DSYRK using $L_{B,J_1}$.
Secondly, it updates the lower part~$L_{B',B}$ of supernode~$J_3$ by invoking DGEMM using $L_{B',J_1}$ and $L_{B,J_1}$.
Finally, it updates the diagonal part~$L_{B',B'}$ of supernode~$J_6$ by invoking DSYRK using $L_{B',J_1}$.


The number of BLAS calls that RLB invokes greatly influences its performance; i.e.,
it should perform better if the blocks are fewer and larger.
This can be done by reordering the columns within the supernodes, which will not affect the amount of fill in the Cholesky factor~\cite{JNP18,KNP24}.

\vspace{-1ex}
\section{GPU Acceleration}\label{sec:factorization}
In both RL and RLB, we utilize GPU for the floating-point computations.
That is, we offload (some) BLAS calls to the GPU.
In order to do this, we add data transfers and replace BLAS calls with their GPU versions.

In the GPU-accelerated version of RL, we add three data transfers.
The first one is to transfer the supernode $J$ to the GPU just before calling DPOTRF, which factors the diagonal block of supernode $J$.
Then, supernode $J$ is factorized on the GPU by calling DTRSM.
After the factorization, we initiate the transfer of supernode $J$ from GPU to CPU.
Here, note that this second transfer is asynchronous since the CPU does not immediately require the data.
Then DSYRK is called to compute the updates from supernode~$J$ for the remaining submatrix.
Note that DSYRK is also called on GPU.
Finally, we transfer the update matrix for supernode $J$ from GPU to CPU.
The rest of the algorithm (i.e., assembly) is done in the CPU.
We also parallelize these assembly loops with OpenMP.

We develop two versions of GPU-accelerated RLB. 
Transferring the current supernode $J$ from CPU to GPU and getting it back from GPU to CPU is the
same as it was in RL for both versions.
Recall that RLB performs updates from supernode~$J$ with individual DSYRK or DGEMM calls for each block.
In the first version, we keep these small update matrices on the GPU until all of them are computed.
Then, when computing the updates for the current supernode $J$ is done,
we transfer all update matrices from GPU to CPU with a single transfer operation.
Then, the updates are assembled in the CPU. 
As we did for RL, we use OpenMP directives here as well.

In the original CPU-only version of RLB, there were no assembly operations;
i.e., updates were directly applied to the factor matrix~$L$ by DSYRK and DGEMM.
However, this nice feature of RLB is not beneficial for the GPU version, since data transfers between GPU and CPU are slow.
In order to use this feature, we need to transfer current values of the ancestor supernodes and then transfer back the updated values.
RLB, in which this feature is not utilized, actually becomes very similar to the GPU version of RL.
The only difference is that the RL update matrix is computed by one large DSYRK call, whereas in RLB it is computed by multiple small DSYRK and DGEMM calls. 
In fact, here RL has the advantage of easier parallelization of one coarse grain task.
Therefore, this version of RLB is of no practical value compared to RL.

In the second version, we transfer back each small update matrix from GPU to CPU as soon as its computation is done.
That is, we perform one transfer and assembly operation for each individual DSYRK or DGEMM call.
The advantage of this version compared to RL and the first version of RLB is lower memory usage. 
In RL and the first version of RLB, the large update matrix must be stored in both CPU and GPU.
This is particularly important since GPU memory is limited.
That is, RL and the first version of RLB cannot be used to factorize certain very large matrices on GPU.

Another optimization we utilized for both RL and RLB is keeping the small computations on CPU.
As we mentioned above, data transfer between CPU and GPU is slow.
Hence, even if the computation in the GPU is much faster, for small volume of data the total transfer and computation time on GPU becomes more than the computation time on CPU.
Therefore, for each supernode we check its size (i.e., the number of nonzeros) and if it is below a threshold, we keep it and all the computation associated with it on CPU.

\vspace{-1ex}
\section{Testing the factorization methods}
\label{sec:tests}
\vspace{-1ex}
\subsection{Setup}
\vspace{-.5ex}
In our experiments, we selected symmetric matrices from the SuiteSparse matrix collection~\cite{DH11}, for which $n \!\geq\!$ \num{600000}.
We excluded matrices that are not realistic sparse linear systems.
We also excluded matrices that require too much time and storage for factorization.
Our dataset contains 21 matrices.

For the ordering step of the solution process, we used metis~\cite{KK99} invoking its nested dissection routine, which lowers the fill in the factor matrix.
In addition, we improved the performance of Cholesky factorization by combining supernodes and applying partition refinement, as described below.

As usual, first the supernode partition was computed using the sparsity structure of $A$~\cite{LNP93}.
However, in general, since the supernodes at the bottom of the supernodal elimination trees are quite small (in terms of the number of columns) and sparse, the work involved is not significant.
Therefore, in order to improve the performance of the factorization, Ashcraft and Grimes~\cite{AG89} proposed merging supernodes so that the partition becomes coarsened.
Today this idea is used in sparse symmetric factorization software packages such as the MA87 package~\cite{HRS10}
and the MA57 package~\cite{Duff04} by default.

We also utilized supernode merging as follows.
We merged supernode pairs~$J$ and $p(J)$ in a sequence, and merging each pair generally increases the fill.
We selected pairs to be merged to minimize at each step the amount of new fill in the factor matrix.
Then our algorithm stopped when the cumulative increase in factor matrix storage went beyond 25\%.

After the coarsening of the supernode partition, we used partition refinement (PR) to reorder the columns of supernodes so that the number of blocks was reduced~\cite{JNP18,KNP24}.
As we mentioned, RLB invokes BLAS calls for the individual blocks in the supernodes.
Therefore this reordering is essential to attain high performance using RLB.
We applied the same PR reorderings for RL as well for consistency.

We ran the experiments on a node of the Perlmutter supercomputer at NERSC,  which has two AMD EPYC 7763 processors (2.45GHz CPUs) 
with~64 cores per socket (128~cores total) and 256~GB of memory, and four Nvidia A100 GPUs having 40~GB of memory.
We compiled our Fortran code with the gfortran 11.2 compiler
using the optimization flag~-O3.

We performed runs where Intel's MKL 2023.2 multithreaded BLAS were linked for CPU calls.
We used MAGMA~\cite{magma} 2.7.1 and Cuda 12.2 for GPU BLAS and data transfer calls, respectively.
We ran the experiments with OpenMP affinity enabled.

\vspace{-1ex}
\subsection{Results}
\label{sec:results}
\vspace{-1ex}

We compared our GPU-accelerated RL and RLB algorithms against their CPU-only versions.
For the CPU runtimes we ran each method using 8, 16, 32, 64, and 128 threads used by the MKL library and used their best.
We also take the best of the RL and RLB runtimes.
In the rest of the paper all speedup values will be given with respect to these ``best'' times. 


For all three GPU accelerated RL and RLB methods, we first ran their GPU only versions. 
Note that here GPU only refers to running all BLAS calls on GPU.
Since data transfer between GPU and CPU is slow as we mentioned, these versions did not achieve reasonable speedup.
In fact, their runtimes were more than CPU-only runtimes for most of the matrices.
Still GPU accelerated RL achieved 3.11$\times$, 3.69$\times$, and 4.15$\times$ speedup on the Long\_Coup\_dt0, Cube\_Coup\_dt0, and Queen\_4147 matrices, respectively.
Note that these are the larger matrices in our test set.
The first version of RLB, which uses the full update matrix, achieved 2.97$\times$ speedup on Queen\_4147, whereas its second version, which uses multiple update matrices, achieved 2.66$\times$ speedup on Queen\_4147.

Then, we ran their second versions, which perform computations associated with large supernodes on GPU and small supernodes on CPU. 
For RL we determined empirically a supernode size threshold of \num{600000}.
That is, if the supernode size (i.e., the number of columns in the supernode times the length of the supernode) is below \num{600000}, then computations associated with this supernode are kept on CPU.
Table~\ref{tab:RLsu} shows runtimes for RL and corresponding speedup values. 
The table also shows the number of supernodes computed on GPU as well as the total number of supernodes.
Note that nlpkkt120 could not be run because its largest update matrix is too big to store on GPU.
As seen in the table, now RL achieves a speedup for each matrix.
The minimum speedup value is 1.31$\times$ achieved on Flan\_1565, whereas the maximum speedup value is 4.47$\times$ achieved on Bump\_2911.
In this setup, the number of supernodes computed on GPU is quite low.

\begin{table}[t]
\scriptsize
\vspace{-1ex}
\caption{Runtimes for GPU accelerated RL together with the speedups and numbers of supernodes computed on GPU}
\vspace{-2.5ex}
\label{tab:RLsu}
\begin{center}
\begin{tabular}{|l||r|r|r|r|} \hline
\multicolumn{1}{|c||}{} &
\multicolumn{1}{c|}{runtime} &
\multicolumn{1}{c|}{ } &
\multicolumn{2}{c|}{\# of supernodes }\\ \cline{4-5}
\multicolumn{1}{|c||}{Matrices} &
\multicolumn{1}{c|}{ (s)} &
\multicolumn{1}{c|}{ speedup} &
\multicolumn{1}{c|}{ on GPU}&
\multicolumn{1}{c|}{ total}\\ \hline
CurlCurl\_2 & 3.800 & 1.59 & 98 & 8,822 \\
dielFilterV2real & 5.599 & 1.40 & 150 & 11,292 \\
dielFilterV3real & 5.669 & 1.43 & 148 & 10,156 \\
PFlow\_742 & 4.497 & 1.35 & 123 & 61,809 \\
CurlCurl\_3 & 7.040 & 2.01 & 164 & 10,074 \\
StocF-1465 & 9.379 & 1.87 & 236 & 40,255 \\
bone010 & 9.158 & 1.41 & 264 & 4,017 \\
Flan\_1565 & 12.853 & 1.31 & 461 & 7,591 \\
audikw\_1 & 9.922 & 1.68 & 264 & 3,725 \\
Fault\_639 & 8.188 & 1.90 & 261 & 1,981 \\
Hook\_1498 & 12.032 & 2.29 & 284 & 10,781 \\
Emilia\_923 & 12.432 & 2.04 & 405 & 2,815 \\
CurlCurl\_4 & 15.745 & 2.44 & 340 & 17,660 \\
nlpkkt80 & 12.596 & 2.42 & 235 & 5,431 \\
Geo\_1438 & 18.698 & 2.01 & 601 & 4,419 \\
Serena & 19.333 & 3.00 & 388 & 4,822 \\
Long\_Coup\_dt0 & 27.708 & 3.22 & 1,432 & 2,897 \\
Cube\_Coup\_dt0 & 42.188 & 3.75 & 2,142 & 3,853 \\
Bump\_2911 & 64.339 & 4.47 & 2,848 & 64,995 \\
nlpkkt120 & & & & 12,785 \\
Queen\_4147 & 89.552 & 4.27 & 3,898 & 7,158 \\ \hline
\end{tabular}
\end{center}
\vspace{-6.5ex}
\end{table}

\begin{figure}[!b]
\vspace{-4ex}
\begin{center}
\includegraphics[width=.8\linewidth]{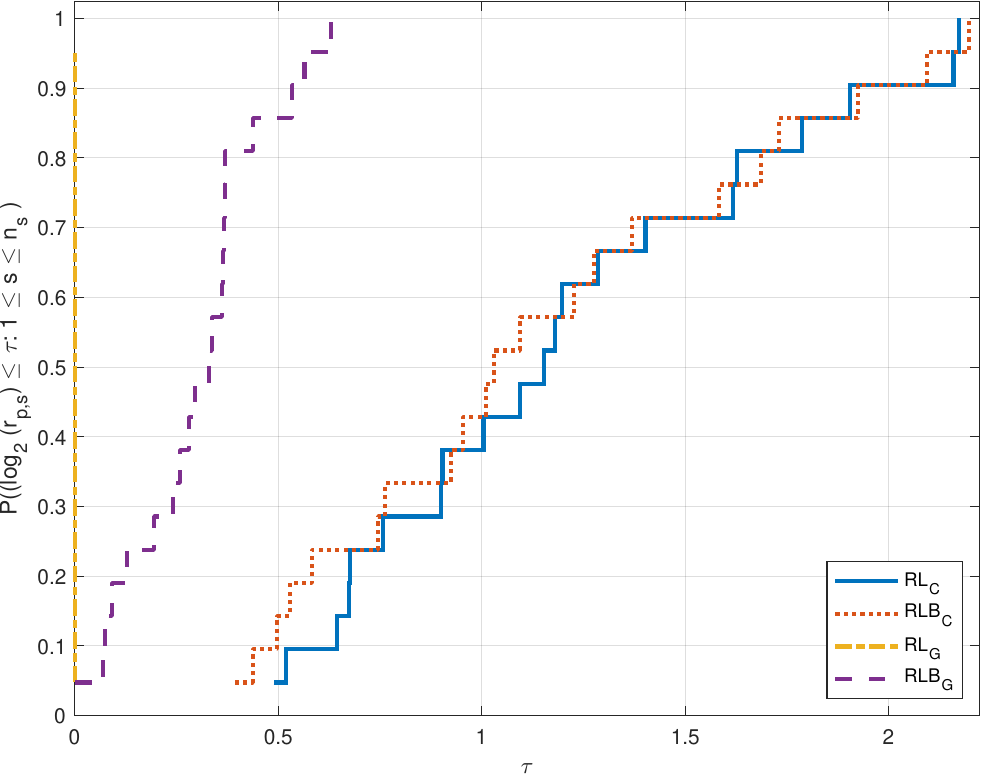}
\vspace{-2ex}
\caption{Performance profile for the factorization times for both CPU and GPU methods. Subscripts ``C'' and ``G'' respectively denote CPU and GPU versions of the RL and RLB methods.}
\label{fig:pp}
\end{center}
\vspace{-3ex}
\end{figure}

As we mentioned, the first version of RLB, which transfers a single update matrix, has no advantage over RL. 
It has the disadvantage of trying to parallelize fine-grain tasks. 
Therefore we do not give any results about this version of RLB.

The second version of the RLB has the advantage of small memory usage compared to RL. 
Table~\ref{tab:RLBsu} shows runtimes for RLB and the corresponding speedup values. 
The table also shows the number of supernodes computed on GPU as well as the total number of supernodes. 
For RLB we determined empirically a supernode size threshold of \num{750000}.
Note that RLB successfully computed the factorization for nlpkkt120.
As seen in the table, also RLB achieves a speedup for each matrix.
The minimum speedup value is 1.09$\times$ achieved on dielFilterV2real, whereas the maximum speedup value is 3.15$\times$ achieved on Queen\_4147.
As expected, the GPU accelerated version of RLB is slower than RL but it can factorize larger matrices.

\begin{table}[t]
\scriptsize
\vspace{-2ex}
\caption{Runtimes for GPU accelerated RLB together with the speedups and numbers of supernodes computed on GPU}
\vspace{-2.5ex}
\label{tab:RLBsu}
\begin{center}
\begin{tabular}{|l||r|r|r|r|} \hline
\multicolumn{1}{|c||}{} &
\multicolumn{1}{c|}{runtime} &
\multicolumn{1}{c|}{ } &
\multicolumn{2}{c|}{\# of supernodes }\\ \cline{4-5}
\multicolumn{1}{|c||}{Matrices} &
\multicolumn{1}{c|}{ (s)} &
\multicolumn{1}{c|}{ speedup} &
\multicolumn{1}{c|}{ on GPU}&
\multicolumn{1}{c|}{ total}\\ \hline
CurlCurl\_2 & 4.802 & 1.26 & 81 & 8,822 \\
dielFilterV2real & 7.204 & 1.09 & 126 & 11,292 \\
dielFilterV3real & 6.776 & 1.20 & 122 & 10,156 \\
PFlow\_742 & 4.715 & 1.29 & 94 & 61,809 \\
CurlCurl\_3 & 9.040 & 1.56 & 146 & 10,074 \\
StocF-1465 & 12.082 & 1.45 & 199 & 40,255 \\
bone010 & 9.754 & 1.32 & 228 & 4,017 \\
Flan\_1565 & 13.529 & 1.25 & 360 & 7,591 \\
audikw\_1 & 11.355 & 1.46 & 223 & 3,725 \\
Fault\_639 & 9.938 & 1.56 & 178 & 1,981 \\
Hook\_1498 & 15.114 & 1.83 & 242 & 10,781 \\
Emilia\_923 & 15.253 & 1.66 & 267 & 2,815 \\
CurlCurl\_4 & 20.324 & 1.89 & 277 & 17,660 \\
nlpkkt80 & 14.886 & 2.05 & 208 & 5,431 \\
Geo\_1438 & 20.419 & 1.84 & 405 & 4,419 \\
Serena & 24.972 & 2.32 & 302 & 4,822 \\
Long\_Coup\_dt0 & 40.968 & 2.18 & 1,207 & 2,897 \\
Cube\_Coup\_dt0 & 61.064 & 2.59 & 1,918 & 3,853 \\
Bump\_2911 & 99.561 & 2.89 & 2,368 & 64,995 \\
nlpkkt120 & 114.658 & 3.07 & 1,048 & 12,785 \\
Queen\_4147 & 121.299 & 3.15 & 3,647 & 7,158 \\ \hline
\end{tabular}
\end{center}
\vspace{-6.5ex}

\end{table}

Here we should briefly compare the two versions of RLB.
On larger matrices, RLB with a single update matrix is up to 9 percent better than RLB with multiple update matrices whereas on smaller matrices, RLB with multiple update matrices is up to 3 percent better than RLB with a single update matrix.
This finding shows that for data transfer between CPU and GPU the latency is negligible but the bandwidth is important.
This is because transferring the same amount of data in a single transfer operation versus multiple transfer operations does not significantly impact performance.

Performance profiles~\cite{dolan2002benchmarking} show a model performs within a factor of performance (x-axis) relative to the best model in what fraction of test matrices (y-axis).
Figure~\ref{fig:pp} shows the performance profile for both the CPU (denoted by subscript ``C'') and GPU (denoted by subscript ``G'') versions of RL and RLB.
As seen in the figure, the GPU version of RL is unequivocally the best, except for one matrix for which RL cannot compute the factorization. 
RLB closely follows RL.
Both RL and RLB using GPU for the BLAS calls with large data are much better than their CPU-only versions.

%
%
\vspace{-1.5ex}
\section{Conclusion}
\label{sec:conclusion}
\vspace{-1.5ex}

We have introduced GPU accelerated variants of 
serial sparse Cholesky algorithms, RL and~RLB.
We added GPU support by offloading large BLAS calls to GPU.
RL computes the update matrix for a supernode with a single BLAS call, whereas RLB computes the update for a supernode with multiple BLAS calls.
Therefore, RL has the advantage of easier parallelism of one coarse-grain task.
On the other hand, the memory footprint of RLB is much lower since it does not need to store full update matrices on both CPU and GPU.

Experiments show that RL achieves up to 4.47$\times$ speedup compared to CPU-only factorization; RLB achieves up to 3.15$\times$ speedup.
Although RL runs faster, RLB is capable of factorizing very large matrices with GPU support.


\bibliographystyle{IEEEtran}

\bibliography{mlf_paper}

\end{document}